\documentclass[10pt]{article}
\usepackage[utf8]{inputenc}
\usepackage[T1]{fontenc}
\usepackage{amsmath}
\usepackage{amsfonts}
\usepackage{amssymb}
\usepackage[version=4]{mhchem}
\usepackage{stmaryrd}
\usepackage{graphicx}
\usepackage[export]{adjustbox}
\graphicspath{{./images/} }

\title{Homomorphic Encryption\\  of Intuitionistic Logic Proofs \\ and Functional Programs:  \\A Categorical Approach \\ Inspired by Composite-Order Bilinear Groups}

\author{Ben Goertzel  \footnote{SingularityNET, TrueAGI, OpenCog } \\
}

\date{}

\let\svthefootnote\thefootnote
\newcommand\blfootnotetext[1]{%
  \let\thefootnote\relax\footnote{#1}%
  \addtocounter{footnote}{-1}%
  \let\thefootnote\svthefootnote%
}

\let\svfootnotetext\footnotetext
\renewcommand\footnotetext[2][?]{%
  \if\relax#1\relax%
    \ifnum\value{footnote}=0\blfootnotetext{#2}\else\svfootnotetext{#2}\fi%
  \else%
    \if?#1\ifnum\value{footnote}=0\blfootnotetext{#2}\else\svfootnotetext{#2}\fi%
    \else\svfootnotetext[#1]{#2}\fi%
  \fi
}

\DeclareUnicodeCharacter{00D7}{\ifmmode\times\else{$\times$}\fi}

\begin{document}
\maketitle

\begin{abstract}
We present a conceptual framework for extending homomorphic encryption beyond arithmetic or Boolean operations into 
the domain of intuitionistic logic proofs and, by the Curry-Howard correspondence, into the domain of typed functional 
programs. We begin by reviewing well-known homomorphic encryption schemes for arithmetic operations, and then 
discuss the adaptation of similar concepts to support logical inference steps in intuitionistic logic. Key to our construction 
are polynomial functors and Bounded Natural Functors (BNFs), which serve as a categorical substrate on which logic 
formulas and proofs are represented and manipulated. We outline a complexity-theoretic hardness assumption -- the BNF 
Distinguishing Problem, constructed via a reduction from Subgraph Isomorphism, providing a foundation for cryptographic 
security. Finally, we describe how these methods can homomorphically encode the execution of total, dependently typed 
functional programs, and outline strategies for making the approach potentially efficient, including software optimizations 
and hardware acceleration.
\end{abstract}

\setcounter{tocdepth}{2}
\tableofcontents

\section{Introduction, Background and Motivation}

Homomorphic encryption (HE) empowers computations on encrypted data without revealing plaintexts. Initial focus was on arithmetic operations, then extended to Boolean functions, and more recently to specialized logical operations. This paper advances this line of work further into the realm of full intuitionistic logic and consequently into the domain of typed $\lambda$ calculi and functional programming, where proofs correspond to programs.

We start by reviewing Tseng et al.'s (2017) homomorphic encryption scheme, which uses composite-order bilinear groups to support logical operations like AND and OR. This scheme encodes bits via subgroup membership, providing a direct logical homomorphism. We then generalize the core concept of this paper to intuitionistic logic using polynomial functors and BNFs -- tools from category theory that allow complex logical structures and infinite data types to be encoded algebraically.

Intuitionistic logic has a well-understood computational interpretation. By the Curry-Howard correspondence, intuitionistic proofs correspond directly to typed $\lambda$-terms in a suitably rich, constructive type theory. Thus, enabling homomorphic computations on intuitionistic proofs also allows for homomorphic evaluation of functional programs (under appropriate constraints like totality and strong normalization).

This paper outlines: (1) state-of-the-art arithmetic and Boolean homomorphic encryption, (2) how to homomorphically handle intuitionistic logic proofs using polynomial functors and BNFs, (3) a complexity-theoretic hardness assumption built from Subgraph Isomorphism, (4) the connection to homomorphic execution of typed functional programs, and (5) potential optimization strategies for plausible performance.

\section{Survey of Existing Homomorphic Encryption Schemes}
\subsection{Arithmetic Homomorphic Encryption}
Early homomorphic encryption schemes aimed at simple algebraic homomorphisms. For instance, the Paillier cryptosystem provides additively homomorphic encryption: given ciphertexts $c_{1}, c_{2}$ encrypting messages $m_{1}, m_{2}$, one can obtain an encryption of $m_{1}+m_{2}$ by multiplying $c_{1} \cdot c_{2}$. Similarly, ElGamal encryption is multiplicatively homomorphic. Combining these in a single system that supports arbitrary polynomial computations without leakage led to the first somewhat and then fully homomorphic encryption (FHE) schemes,\\
notably Gentry's breakthrough construction, and subsequent lattice-based schemes (BGV, BFV, CKKS).

These arithmetic-focused schemes generally work over rings or ideal lattices, with security based on problems like Learning With Errors (LWE). They allow complex arithmetic computations but do not directly support logical inference or richer combinational structures unless those are compiled down to arithmetic circuits at great cost in complexity.

\subsection{Extending Homomorphisms to Logical Operations Using Composite-Order Groups}
The path we outline toward homomorphic encryption of logic begins with the rather restricted-scope ideas of Tseng et al. (2017), who described a scheme specifically designed for logical operations -- AND and OR -- in contrast to the prevalent focus on arithmetic. While arithmetic HE relies on adding and multiplying ciphertexts, Tseng et al.'s approach uses composite-order bilinear groups to encode bits as elements of distinct subgroups, thus achieving a direct "logical" homomorphism.

\subsubsection{Overview of Tseng et al.'s Scheme}

The scheme sets up a bilinear group $G$ of composite order $N=p q$, where $p$ and $q$ are large primes. Within this group, there exist subgroups $G_{p}$ and $G_{q}$ of order $p$ and $q$, respectively.

A bilinear map $e: G \times G \rightarrow G_{T}$ satisfies bilinearity, non-degeneracy, and subgroup cancellation properties.

\paragraph{Representation of Bits via Subgroup Membership:}

Each bit $m \in\{0,1\}$ is encrypted as an element of one of the subgroups:

\begin{itemize}
  \item If $m=1$, the ciphertext resides in $G_{p}$.
  \item If $m=0$, the ciphertext is in $G_{q}$.
\end{itemize}

This encoding leverages the hardness of distinguishing subgroup elements in composite-order groups. Without knowing the factorization of $N$ (and thus not knowing how to separate $G_{p}$ and $G_{q}$ ), an adversary cannot determine the plaintext bit.

\subsubsection{Homomorphic OR and AND:}
The scheme supports logical OR and AND as follows:

\begin{itemize}
  \item OR: Given two ciphertexts $c_{1}$ and $c_{2}$ encoding bits $m_{1}$ and $m_{2}$, the ciphertext for $m_{1} \vee m_{2}$ is obtained by simply multiplying $c_{1} \cdot c_{2}$ in $G$. Due to the structure of $G$, if either $c_{1}$ or $c_{2}$ originated in $G_{p}$, the product lies in a subgroup that decrypts to 1.
  \item AND: To perform logical AND, the scheme uses the bilinear pairing $e$. Pairing two ciphertexts $c_{1}, c_{2}$ that encrypt $m_{1}, m_{2}$ yields $e\left(c_{1}, c_{2}\right)$, which represents $m_{1} \wedge m_{2}$. If both are from $G_{p}$, the pairing reflects a logical AND that decrypts to 1 ; otherwise, it decrypts to 0 .
\end{itemize}

\subsubsection{Security and Correctness}
The security relies on the subgroup decision assumption: distinguishing whether a random element lies in $G_{p}$ or $G_{q}$ is hard. Thus, the bit remains secret. Correctness follows from the group-theoretic properties ensuring that performing OR or AND on ciphertexts corresponds to the intended logical operation on the underlying bits.

\subsubsection{Application to 2-DNF and k-CNF}
Tseng et al. extended their scheme to evaluate 2-DNF and k-CNF formulas. By recursively combining ORs and ANDs, they showed that complex Boolean formulas can be evaluated homomorphically, with ciphertexts corresponding to the final outcome.

\subsubsection{Limitations and Extensions}
While this scheme significantly advances beyond arithmetic or simple Boolean gates, it remains focused on propositional formulas. Our goal is to push these ideas further-to entire intuitionistic logic proofs and, by extension, typed functional programs. The Tseng et al. scheme inspires the next steps: replacing "modding by integers" with "quotienting by BNFs" to handle more complex logical constructs.

\section{Intuitionistic Logic, Polynomial Functors, and Bounded Natural Functors: Some Categorical and Algebraic Details}

To homomorphically encrypt intuitionistic logic proofs, we must encode logical formulas and their proofs into algebraic structures that support well-defined transformations corresponding to inference steps.  For this we require a setting in which logical connectives and quantifiers correspond to stable, compositional algebraic operations. Polynomial functors and BNFs offer such a setting, which we now summarize.

\subsection{Intuitionistic Logic and Categorical Semantics}
Intuitionistic logic can be presented as a type theory, where propositions correspond to types and proofs correspond to terms inhabiting these types. Each logical connective-such as implication $(\rightarrow)$, conjunction $(\wedge)$, disjunction $(\vee)$, and quantifiers $(\forall, \exists)$-has a well-understood interpretation in constructive type theory. The semantics of intuitionistic logic can be given in a topos or a category with suitable structure, where objects represent propositions/types and morphisms represent proofs.

\subsection{Polynomial Functors}
A polynomial functor is an endofunctor on Set (or another base category) built from basic operations: products, coproducts, and exponentials of the identity functor. Formally, a finitary polynomial functor $F$ : Set $\rightarrow$ Set can be written as

$$
F(X)=\sum_{i \in I} X^{J_{i}}
$$

\noindent for some indexing sets $I, J_{i}$. Intuitively, $F$ behaves like a "polynomial" in the variable $X$. Such functors naturally represent algebraic data types. For example, a binary tree functor can be expressed as a polynomial functor, and so can finite structures like lists. In logic, these functors capture the shape of propositions. For instance, implication types can be represented using exponentials, while product types represent conjunction and coproduct types represent disjunction.

\subsection{Bounded Natural Functors (BNFs)}
A Bounded Natural Functor (BNF) is a polynomial-like functor equipped with additional structural properties ensuring compositional well-behavior. BNFs often arise in the semantics of complex type constructors like inductive and coinductive types. By choosing a family $\mathcal{B}$ of BNFs with certain closure and boundedness conditions, we can form quotients and subfunctors that preserve the categorical structure needed to represent logical inference steps.

BNFs provide a smooth setting for "modding out" parts of structures. Just as arithmetic schemes use "mod $N$ " to hide details, here we use a "quotient by $\Phi^{\prime \prime}$ for some $\Phi \in \mathcal{B}$ to hide structural details of a proof.

\subsection{Representing Intuitionistic Proofs as Functorial Data}
Every intuitionistic formula and proof can be encoded as an object in a category $\mathbf{C}$ of polynomial (or BNF-based) structures. For example:

\begin{itemize}
\item {\bf Propositions as Functors:} A simple proposition like $A \wedge B$ can be represented by the product of functors representing $A$ and $B$. Implication $A \rightarrow B$ corresponds to an exponential functor. 
\item More complex constructions (like $\forall x . P(x)$ ) can be represented using polynomial functors that model potentially infinite families of instances (via fixpoints or folds).
\end{itemize}

\section{Homomorphic Encryption via Representing Proofs as Morphisms:}

Given this setup, we can now conveniently represent proofs as morphisms, which allows us to naturally encrypt them using BNFs following the conceptual pattern from Tseng et al's arithmetic HE scheme.

The starting-point here is the observation that: A proof of $C$ from $\Gamma=\left\{A_{1}, \ldots, A_{n}\right\}$ corresponds to a morphism between objects (functors) representing $\Gamma$ and $C$. Logical inference steps correspond to natural transformations and compositions of functors.

\subsection{Quotienting by BNFs to Encrypt}
To encrypt a proof represented by a functor $F$, we choose a random BNF $\Phi$ and form the quotient $F / G_{\Phi}$, where $G_{\Phi}$ is a subfunctor (or equivalence relation) defined by $\Phi$. This constructs a new functor $F / G_{\Phi}$ in which certain distinctions are "collapsed." Without knowledge of $\Phi$, the structure $F / G_{\Phi}$ looks like a scrambled version of $F$.

Just as Tseng et al.'s scheme used group substructures to hide bits, here we use BNF-induced equivalences to hide the underlying structure of proofs. The notion of "encrypting" a proof thus becomes "selecting a random quotient by a BNF," rendering the proof's exact identity indistinguishable.

\subsection{Homomorphic Inference Steps}
Logical inference rules -- introduction and elimination rules for connectives, or quantifier instantiations -- translate into algebraic operations on functors. Since we have a quotient structure, these operations lift to the quotient. Thus, performing an inference step on encrypted proofs corresponds to applying a natural transformation or composition on the quotient. The homomorphic property ensures that these encrypted operations correspond to the same logical steps on the underlying, original proofs.

\subsection{Handling Quantifiers and Infinite Structures}
Quantifiers introduce complexity. Universal quantification $\forall x . P(x)$ corresponds to proving $P(x)$ for all $x$. Representing such a proof might involve a functor that encodes a "function" from each element $x$ to a proof of $P(x)$. Polynomial functors can encode infinite data types via folds or fixpoints, and BNFs support forming quotients of these folded structures. Although more complex, the theory of BNFs (as discussed in Avigad et al., Furer et al.) gives tools to handle these situations: one can represent infinite constructions and still define well-behaved quotients.

\subsection{Theoretical Coherence and Security Implications}
By carefully choosing a large family $\mathcal{B}$ of BNFs and randomly selecting $\Phi$ from it, we ensure that decrypting (inverting the quotient) is as hard as solving a specified algorithmic problem we refer to as the "BNF Distinguishing Problem."  This parallel to the subgroup decision assumption in Tseng et al.'s scheme gives us a cryptographic reason to trust that the original structure (the proof) remains hidden.

In summary, the combination of polynomial functors for representing logical data and BNFs for constructing cryptographically hard quotients forms the heart of our approach, extending the homomorphic encryption paradigm from mere bits and Boolean formulas to full-blown intuitionistic logic proofs.

\section{Establishing the Complexity of the BNF Distinguishing Problem}
To claim that the scheme  we have sketched is actually secure, we need a hardness assumption ensuring that given a quotient $F / G_{\Phi}$, it is difficult to recover $\Phi$ or distinguish between different BNFs. We refer to this cryptographic challenge as the "BNF Distinguishing Problem."  Here, we detail the formulation of the BNF Distinguishing Problem, and how it relates to a known, better-understood hard problem: Subgraph Isomorphism (GI).

\subsection{Defining the BNF Distinguishing Problem}
Let $\mathcal{B}$ be a large family of BNFs, each representable by a finite structure (like a syntax tree of sum/product/exponential constructions, possibly with fixpoints). Suppose we have a universal polynomial functor $F$ capturing the structure of interest (e.g. a "universal" encoding of all relevant proposition forms).

For each $\Phi \in \mathcal{B}$, we define a quotient $F / G_{\Phi}$. Encryption corresponds to picking a random $\Phi$ , computing $F / G_{\Phi}$, and giving the result to an adversary. The BNF Distinguishing Problem is: given $F / G_{\Phi}$ (but not $\Phi$ ), can the adversary identify $\Phi$ or even distinguish it from another candidate $\Phi^{\prime}$ ?

Intuitively, if $\mathcal{B}$ is large and the structure of $F / G_{\Phi}$ appears "random," inverting the quotient should be as hard as some known difficult computational problem. We propose Subgraph Isomorphism as a source of such hardness.

\paragraph{The Subgraph Isomorphism Problem} The Subgraph Isomorphism (GI) problem asks: given two finite graphs $G=\left(V_{G}, E_{G}\right)$ and $H=$ $\left(V_{H}, E_{H}\right)$, determine whether  G is isomorphic to a subgraph of H. This problem is known to be NP-complete for general graphs, and is widely considered to be difficult to solve in practice.

\subsection{Encoding Subgraph Isomorphism Using BNFs}

A sketch of how one might encode subgraph isomorphism into a BNF-based construction would look like:

\begin{enumerate}
  \item Define a "universal" polynomial functor $U$ that encodes a broad class of graph-like structures.
  \item For each pattern (the "small" graph $P$ ) and each larger "host" structure $H$, define a corresponding BNF $\Phi_{P}$ that enforces identifications reflecting the presence of $P$ as a subgraph of $H$.
  \item Show that distinguishing which BNF $\Phi_{P}$ was used to form a quotient of $U$ (and possibly combined with information derived from $H$ ) reduces to deciding whether $P$ is embeddable as a subgraph of $H$.
\end{enumerate}

\noindent If successful, this construction means that breaking the BNF-based encryption (i.e., distinguishing which quotient you have) is at least as hard as subgraph isomorphism.

We now walk through these steps in moderately more detail.

\subsubsection{Step-by-Step Outline}

\paragraph{STEP 1: Representing Graphs as Polynomial Functors}.  Consider how to represent a graph as a functor. For a finite graph $G=(V, E)$, we can associate to it a polynomial functor $F_{G}$ : Set $\rightarrow$ Set that, given a set $X$, returns some structured data encoding a labeling of $V$ and adjacency information encoded in a systematic way.

For instance:

\begin{itemize}
  \item Vertices might be represented as indices in a product.
  \item Edges might be represented using factors that "select" pairs of elements corresponding to edges.
\end{itemize}

A simple example: if you just want to represent "a labeled graph structure" as data, you could define

$$
U(X)=\prod_{v \in V_{H}} X^{\operatorname{deg}(v)}
$$

\noindent where $\operatorname{deg}(v)$ is the degree of vertex $v$ in the host graph $H$, and each factor $X^{\operatorname{deg}(v)}$ encodes the choices of neighbors. This is a very rough idea; more sophisticated encodings can capture the full adjacency structure. The key idea is that $U$ is a polynomial functor whose shape depends on a large "universal" indexing construction that can represent many graphs and their potential subgraphs.  (NOTE: This might relate to MORK somehow, but that's an even more speculative aside to the current considerations... though it could conceivably end up relevant to implementing this stuff should we ever get there...)

\paragraph{Universal Functor $U$}. Ideally, $U$ is chosen to be very general, able to encode not just one fixed graph but a whole range of graph-like configurations. It might be parameterized by sets that represent the nodes and edges. This makes $U$ a "template" functor from which we can specialize to represent any host graph $H$.

\paragraph{STEP 2: Representing the Pattern $P$}

\paragraph{Pattern Graph $P$}. Suppose we have a small pattern graph $P=\left(V_{P}, E_{P}\right)$. We want to know if $P$ is a subgraph of $H=\left(V_{H}, E_{H}\right)$.

\paragraph{ $\quad$ BNF $\Phi_{P}$ }  A Bounded Natural Functor (BNF) $\Phi_{P}$ can be constructed to represent the "pattern constraints" of $P$. Intuitively, $\Phi_{P}$ encodes how we would identify certain parts of a larger graph structure if it contained a subgraph isomorphic to $P$.

For example:

\begin{itemize}
  \item Consider that an isomorphic embedding of $P$ into $H$ is a mapping $\pi: V_{P} \rightarrow V_{H}$ preserving edges.
  \item We can use a BNF that, given a large structure encoded by $U(X)$, identifies those components corresponding to vertices and edges of $P$ with a specific pattern inside $U(X)$. If $P$ can be embedded in $H$, then these identifications can be "resolved" to produce a simplified quotient structure with a recognizable normal form.
\end{itemize}

More concretely, $\Phi_{P}$ might be defined so that it imposes equivalence relations on certain indices of $U$ (corresponding to would-be vertices and edges of $P$ ) to reflect the required adjacency pattern. If $P$ is actually present as a subgraph of $H$, these identifications collapse a part of $U(X)$ into a form that can only occur if the subgraph is there.

\paragraph{STEP 3: Quotienting by the BNF $\Phi_{P}$}

\paragraph{ Forming the Quotient $U / G_{\Phi_{P}}$}. Applying the chosen BNF $\Phi_{P}$ leads to a quotient of $U$ by a certain equivalence relation induced by $\Phi_{P}$. This quotient can be thought of as "enforcing the pattern $P$ " on the universal structure.

We have the cases:

\begin{itemize}
\item {\bf If $P$ is a Subgraph of $H$ :} Then the quotient $U / G_{\Phi_{P}}$ (when specialized to the structure representing $H$ ) will have a particular canonical form or normal shape. Essentially, the identifications required by $\Phi_{P}$ line up perfectly with a subset of $H$ 's structure, producing a neat factorization or collapsing of $U(H)$.
\item {\bf If $P$ is not a Subgraph of $H$ :}. The imposed identifications won't match any part of the structure representing $H$. As a result, the quotient $U / G_{\Phi_{P}}$ might look "random" or fail to produce the same canonical form. The absence of a neat matching pattern means the quotient structure differs in a detectable way.
\end{itemize}

\paragraph{STEP 4: Distinguishing BNFs Encodes Subgraph Isomorphism}

\paragraph{Family of BNFs for Different Patterns:}. Now consider a family of BNFs $\left\{\Phi_{P_{i}}\right\}$, each corresponding to a different pattern graph $P_{i}$. Given a host graph $H$, if we apply one of these BNFs and get a quotient structure, distinguishing which $\Phi_{P_{i}}$ was used is equivalent to identifying which pattern $P_{i}$ occurs as a subgraph of $H$.

\paragraph{ Reduction to Subgraph Isomorphism:}  If we had an efficient procedure for distinguishing which BNF induced the given quotient, we could solve subgraph isomorphism. We'd pick a pattern $P$, construct $\Phi_{P}$, compute $U / G_{\Phi_{P}}$ relative to $H$, and then check if the resulting structure corresponds to the "subgraph-present" canonical form. Being able to distinguish the BNF means we can distinguish whether $P$ appears as a subgraph in $H$.

\subsubsection{Discussion}

The above procedure leverages the fact that BNFs are not just simple polynomials; they can represent complex, possibly infinite, type constructors. They allow the encoding of intricate equivalence relations and identifications. By appropriately designing $\Phi_{P}$, we impose exactly the structural constraints that correspond to the presence of $P$ inside $H$.

The complexity of subgraph matching -- an inherently combinatorial problem -- can be represented as finding a suitable functorial pattern. The BNF's role is to "mask" or "unmask" structural details that correspond to a pattern. Distinguishing different BNFs then corresponds to identifying whether certain structural patterns exist, i.e., solving subgraph isomorphism.

Unlike polynomial factorization over the integers or rationals, here the complexity is combinatorial and structural. The BNF-quotienting process does not reduce easily to a known, efficiently solvable algebraic factorization problem. Instead, it maps to an NP-complete pattern detection problem, based on subgraph isomorphism.

\subsubsection{Re-cap}

In sum: By constructing a universal functor $U$ that can represent a large class of graphs and defining BNFs $\Phi_{P}$ that impose identifications corresponding to subgraph patterns, we can ensure that distinguishing which BNF (and thus which pattern) is encoded in a quotient is at least as hard as the subgraph isomorphism problem. In this way we see that the complexity of the BNF Distinguishing Problem can be anchored in a well-known NP-complete problem, providing a reasonably sound basis for the claim that distinguishing BNFs is computationally intractable.

\section{From Intuitionistic Proofs to Program Execution}
The Curry-Howard correspondence links intuitionistic logic proofs and typed $\lambda$-terms:

\begin{itemize}
  \item Proofs $\leftrightarrow$ Programs: Each proof of a proposition is a term of a corresponding type.
  \item Inference Steps $\leftrightarrow$ Evaluations: Logical inference corresponds to computation steps, such as $\beta$-reduction in the $\lambda$-calculus.
\end{itemize}

Thus, homomorphic encryption of logic proofs naturally extends to homomorphic encryption of program execution. If we can evaluate inference steps on encrypted proofs, we can similarly perform computations on encrypted programs.

To ensure that program evaluation corresponds neatly to proof normalization, we focus on languages that are strongly normalizing, ensuring every program eventually terminates. Dependently typed languages like Idris or Agda can enforce totality; as can languages like MeTTa with extremely flexible type system support.  By restricting attention to total functions and well-founded recursive definitions, these languages guarantee that all programs terminate and have a direct constructive interpretation as proofs in intuitionistic logic.

\subsection{The Example of Idris}
Idris (Brady, 2013) is a dependently typed language designed with an emphasis on totality checking and practical programming. Through its dependent type system, one can ensure that all defined functions and data structures are total, meaning they always terminate and yield a value. Under the Curry-Howard view, these total functions correspond to fully constructive, normalizing proofs.

By working in a fragment of Idris where totality is enforced, we get:

\begin{itemize}
  \item Strong Normalization: Every Idris term in this fragment corresponds to a terminating proof.
  \item Constructive Semantics: Operations in Idris align closely with intuitionistic logic inference rules.
  \item No Non-Termination Worries: Evaluating an encrypted Idris program homomorphically will always yield a result, ensuring that the homomorphic encryption process for proofs translates cleanly into homomorphic program evaluation.
\end{itemize}

\subsubsection{Homomorphic Execution of Idris Programs}. 

The application of our homomorphic encryption scheme to Idris programs is straightforward given the machinery set up.

If we encrypt the "proof" (i.e., the Idris program) by selecting a random BNF and quotienting the universal functor, the server holding the ciphertext can apply the corresponding evaluation steps—encoded as functorial transformations-without ever seeing the plaintext. Since every evaluation step corresponds to a logical inference, each step is homomorphically executable.

The client, knowing the chosen BNF, can decrypt the final normal form and thus recover the computed result. In essence, homomorphic evaluation of total Idris programs is homomorphic normalization of intuitionistic proofs, realized via the categorical and functorial constructs.

\subsubsection{Practical Implications and Restrictions}
While this approach is elegant, it demands restricting to a total, strongly normalizing subset of Idris.  If one is willing to make this restriction, however, one obtains direct and conceptually sound environment for applying our homomorphic techniques. This yields a scenario where one can, in principle, outsource arbitrary provably terminating computations to an untrusted party, run them homomorphically on encrypted inputs, and retrieve encrypted outputs, all guaranteed to be correct under intuitionistic logic interpretations.

\subsection{The Example of MeTTa}

The MeTTa AI scripting language (Meredith et al, 2023), while less mature than Idris, poses another natural candidate for homomorphic encryption of programs according to the present methods.   MeTTa does not require use of types or type systems, but makes it straightforward to create and leverage highly flexible type systems.   One can create type systems possessing desired properties such as totality, and then potentially apply homomorphic encryption methods like the ones outlined here to MeTTa programs constructed using these given types.   

For some AI applications, it is more convenient to apply MeTTa in a more general sense without these sorts of type-theoretic restrictions.   However, if applying MeTTa as a smart contract language, for example, then it will generally be apropos to apply even more restrictive type systems than needed for homomorphic encryption, so as to make formal verification of smart contract properties tractable and rapid.   Overall there should be a significant variety of MeTTa use cases in which it is not overly restrictive to utilize a type system friendly to the current homomorphic encryption methods.

\section{A Simple Example}

To make the above a little more concrete, in this section we spell out how this would work in a specific "toy scale" example problem:

Our goal in the example problem is: {\it Represent integers modulo 4 and perform addition homomorphically on their encrypted representations.}  

We will encode these integers as certain functions on a four-element set. The "operation" we perform homomorphically is function composition, which will correspond to addition mod 4 on the underlying integers. The encryption will use a BNF-based quotient to hide which functions correspond to which integers.

\subsection{Plaintext Setup}

First let's pose the problem formally in a manner suitable for our homomorphic encryption scheme.

\subsubsection{Representing the Data}
Let $A=\left\{a_{0}, a_{1}, a_{2}, a_{3}\right\}$. Consider the set of all functions $f: A \rightarrow A$. Each such function can be represented by a 4-tuple $\left(f\left(a_{0}\right), f\left(a_{1}\right), f\left(a_{2}\right), f\left(a_{3}\right)\right)$, where each $f\left(a_{i}\right) \in A$.

Since $|A|=4$, there are $4^{4}=256$ possible functions from $A$ to $A$.

\paragraph{Encoding Integers mod 4 as Functions (Plaintext)} We will represent integers $\{0,1,2,3\} \bmod 4$ using "shift functions." Define $f_{k}: A \rightarrow A$ by:

$$
f_{k}\left(a_{i}\right)=a_{(i+k) \bmod 4} .
$$

For example:

\begin{itemize}
  \item $f_{0}$ is the identity: $f_{0}\left(a_{0}\right)=a_{0}, f_{0}\left(a_{1}\right)=a_{1}, f_{0}\left(a_{2}\right)=a_{2}, f_{0}\left(a_{3}\right)=a_{3}$.
  \item $f_{1}$ shifts each element by $1: f_{1}\left(a_{0}\right)=a_{1}, f_{1}\left(a_{1}\right)=a_{2}, f_{1}\left(a_{2}\right)=a_{3}, f_{1}\left(a_{3}\right)=a_{0}$.
  \item $f_{2}$ shifts by 2 , and $f_{3}$ by 3 similarly.
\end{itemize}

\subsubsection{Addition via Composition (Plaintext)}

Function composition, in this setup, corresponds to addition mod 4:

$$
f_{k} \circ f_{\ell}=f_{(k+\ell) \bmod 4} .
$$

\noindent So, composing these shift functions adds their corresponding integers mod 4.

\subsubsection{The Universal Functor}

Let $F(X)=X^{4}$. For any set $X, F(X)$ is the set of 4-tuples of elements from $X$. Since polynomial functors like $X^{4}$ are well-behaved, $F$ is a BNF.

\begin{itemize}
  \item When $X=A, F(A)=A^{4}$ represents all functions $A \rightarrow A$.
  \item Each integer $k \bmod 4$ is represented by a unique tuple in $A^{4}$ (the one induced by $f_{k}$ ).
\end{itemize}

\noindent Without encryption, the evaluator can see exactly which tuple corresponds to which $f_{k}$, and thus knows which integer is encoded.

\subsection{Encryption via a Chosen BNF $\Phi$}

Now let's walk through how the encryption works in this example.

\subsubsection{Introducing a Quotient}

To encrypt, we select a secret BNF $\Phi$ that induces an equivalence relation $R_{\Phi}$ on $F(A)=A^{4}$ . This equivalence relation identifies certain 4-tuples so that the evaluator cannot distinguish which one is $f_{0}, f_{1}, f_{2}$, or $f_{3}$.

For simplicity, suppose $\Phi$ groups the 256 functions into equivalence classes of size 4 , with each class containing exactly one of the shift functions $f_{0}, f_{1}, f_{2}, f_{3}$ along with three other "dummy" functions. A conceptual pattern might be:

\begin{itemize}
  \item One equivalence class contains $\left\{f_{0}, g_{1}, g_{2}, g_{3}\right\}$, four distinct tuples, one of which is the identity shift $f_{0}$.
  \item Another class contains $\left\{f_{1}, h_{1}, h_{2}, h_{3}\right\}$.
  \item Another for $\left\{f_{2}, i_{1}, i_{2}, i_{3}\right\}$, and so on.
  \item Each $f_{k}$ is "hidden" inside a class of four tuples that look unrelated, making it hard to tell which one is the $f_{k}$ function representing $k$.
\end{itemize}

\subsubsection{Quotient Functor:}
Formally, we define:

$$
F_{\Phi}(A)=F(A) / R_{\Phi},
$$

\noindent where $R_{\Phi}$ is the equivalence relation induced by $\Phi$. By the closure properties of BNFs, $F_{\Phi}$ is also a BNF.

\subsubsection{Ciphertexts:}
An encrypted integer $k$ mod 4 is given as a certain equivalence class $\left[f_{k}\right]$ that also includes other, non-shift functions. Without knowing $\Phi$, the evaluator cannot discern which member of the class is the real $f_{k}$.

\subsection{Performing a Computation: Addition via Composition}

Now, finally, let's run through how this encryption scheme works in performing an example computation!

\subsubsection{Public Operations}
The evaluator knows how to perform function composition on ciphertexts at the quotient level. Composition corresponds to a binary operation $\star$ defined on the equivalence classes:

$$
\left[f_{x}\right] \star\left[f_{y}\right]=\left[f_{x} \circ f_{y}\right]
$$

Although the evaluator sees only equivalence classes, the composition rule is publicly defined. For any representatives chosen consistently, there's a well-defined way to compose classes. The evaluator can do this without knowing which class corresponds to which shift $f_{k}$.

\subsubsection{Example Computation}

\paragraph{Inputs} Suppose we have two encrypted integers representing $f_{1}$ and $f_{3}$. The first ciphertext is $\left[f_{1}\right]$, an equivalence class containing $f_{1}$ and some dummy functions. The second ciphertext is $\left[f_{3}\right]$, another class containing $f_{3}$ plus dummies.

\paragraph{Performing Addition Homomorphically} The evaluator applies the known operation $\star$ :

$$
\left[f_{1}\right] \star\left[f_{3}\right]=\left[f_{1} \circ f_{3}\right]
$$

\noindent They know how to combine the classes by following the public "composition on classes" rules (e.g., "compose representatives and then take the quotient"). After this step, they get a new equivalence class $\left[f_{1+3 \bmod 4}\right]=\left[f_{0}\right]$, which encodes the sum $(1+3) \bmod$ $4=0$\\

\paragraph{Evaluator's Perspective}  The evaluator sees:

\begin{itemize}
  \item A ciphertext $\left[f_{1}\right]$ that they know is one of several possible functions.
  \item A ciphertext $\left[f_{3}\right]$ that is similarly ambiguous.
  \item After composition, they get $\left[f_{0}\right]$ which is also ambiguous, containing multiple possible functions.
\end{itemize}

\noindent They have performed a meaningful computation (adding two secret numbers mod 4) entirely in the encrypted domain. They did so by using public rules for composition at the quotient level.

The evaluator does not know which equivalence class corresponds to which integer. They only know that they've combined two ciphertexts according to a known operation. Without knowing $\Phi$, they cannot map classes back to the original $f_{k}$ functions or the integers $k$. Thus, they can carry out the homomorphic computation "blindly."

\paragraph{No Obfuscation!!} To be noted: All the operations here (composition, quotient manipulation) are public and well-defined. There's no code obfuscation. The complexity assumption ensures that reversing the quotient to determine which class corresponds to which $f_{k}$ is hard, just like factoring $n$ in RSA is believed to be hard. The "secret key" is the particular $\Phi$ chosen. Without $\Phi$, the evaluator cannot decrypt or break the scheme.

\section{Performance Considerations and Potential Optimizations}
While the theoretical framework outlined here is (we believe) compelling, naive implementations would be highly inefficient. Operating on polynomial functors and BNFs involves complex algebraic and categorical transformations, dwarfing the cost of simpler arithmetic or Boolean circuits. Achieving practical performance will require a multi-pronged optimization strategy, at which we will now vaguely gesture.

\subsection{Software-Level Optimizations}
\paragraph{Normal Forms and Canonical Representations:}
Converting polynomial functors into canonical normal forms before homomorphic operations can simplify transformations. A normal form reduces the complexity of repeatedly applying compositions, quotients, and natural transformations.

\paragraph{Precomputation and Caching:}
Many functor patterns, especially those recurring frequently in a given workload, can be precomputed and cached. By storing results of common quotient operations or canonical transformations, we can avoid recomputing expensive steps. Similarly, partial evaluations and memoization of subexpressions help tame exponential blowups.

\paragraph{Algebraic Simplifications Using Functor Laws:}
Polynomial functors and BNFs obey algebraic laws (distributive, associative, etc.). Applying these laws to simplify a complex composition or quotient before executing it can drastically reduce computational overhead. Detecting and exploiting isomorphic structures and factorizing equivalences can transform a complex operation into a simpler, more direct one.

\paragraph{Incremental and Lazy Computation} Instead of fully evaluating all transformations upfront, lazy evaluation defers computation until results are actually needed. Incremental updates mean that if only a part of the structure changes, we only recompute that part, not the entire quotient.

\paragraph{Domain-Specific Languages (DSLs) and Compiler Support} A DSL for describing polynomial functor transformations and BNF quotients can feed into a specialized compiler that applies domain-specific optimizations. The compiler could perform strength reduction, loop fusion, or partial evaluation at compile time, generating code specialized for the current functor structure.

\paragraph{Parallelization and Distributed Computation} Polynomial functor manipulations often decompose into subproblems on independent components. By parallelizing these computations across multi-core CPUs, GPUs, or distributed clusters, we can harness concurrency for speedups. For example, separate subfunctors can be processed in parallel, and partial results combined at the end.

\subsection{Hardware Acceleration for Polynomial Functor Operations}
While software optimizations help, the complexity of operations suggests that specialized hardware could provide substantial gains, much as GPU acceleration benefits machine learning workloads.

\subsubsection{Arithmetic Accelerators:}
Polynomial functors often require large integer arithmetic and polynomial multiplications. Hardware acceleration of modular exponentiation, polynomial multiplication (e.g., via FFT-based algorithms implemented in ASICs), and elliptic curve pairings (crucial in Tseng et al.-style schemes) can deliver orders-of-magnitude improvements. Field-programmable gate arrays (FPGAs) or Application-Specific Integrated Circuits (ASICs) can implement these operations more efficiently than a CPU's general-purpose instruction set.

\subsubsection{Domain-Specific Accelerators (DSAs) for Functor Transformations:}
Beyond raw arithmetic speedups, one can design DSAs that directly support common polynomial functor operations. For example:

\begin{itemize}
  \item Functor Composition Units: Specialized circuits to handle sums, products, and exponentials at the hardware level. These units could store representation details of polynomial functors in compressed forms and apply known rewriting rules directly in hardware.
  \item Hardware-Assisted Canonicalization: Implementing logic for canonical form transformations as a microcoded sequence of instructions or even hardwired logic. If certain BNF patterns occur often, lookup tables or hardware caches could quickly recognize and simplify them.
  \item Dedicated Circuits for Quotienting by BNFs: A hardware block could handle equivalence classes induced by BNFs, applying predefined pattern-matching logic to identify and collapse equivalent substructures. As BNFs define structured equivalences, a hardware unit might apply these equivalences via preloaded "pattern tables," significantly reducing the complexity of quotient operations.
\end{itemize}

\subsubsection{Memory and Storage Optimizations:}
Polynomial functor operations may be memory-bound if large data structures move repeatedly between CPU and memory. A hardware accelerator integrated tightly with main memory or using high-bandwidth memory (HBM) could reduce latency. Caches or on-chip SRAM could store commonly accessed patterns, supporting near-data processing and minimizing data movement overhead.

\subsubsection{Instruction Set Extensions:}
Extend the CPU's instruction set with custom instructions for polynomial functor manipulation. Similar to how CPUs have vector instructions (SIMD) for parallel arithmetic, we might add "functor operations" that handle common transformation steps at the instruction level, offloading complexity from software.

\subsection{Balancing Security, Complexity, and Performance}
Practical deployment must balance parameter sizes and complexity. Smaller parameters or restricted classes of BNFs improve performance but may reduce security. Domain-specific accelerators will need to be flexible enough to handle various BNF families and polynomial functor sizes while maintaining security margins.

Ultimately, achieving practical performance for real-world problems will likely involve a combination of all the above strategies: carefully chosen parameters, sophisticated software optimizations, parallelization, and custom hardware acceleration.

\section{Conclusion and Future Work}
We have sketched a broad vision: starting from arithmetic homomorphic encryption, we broaden the scope to encrypting intuitionistic logic proofs homomorphically. By leveraging polynomial functors and BNFs, we generalize beyond Boolean operations to full logical inference steps. Through Curry-Howard, this extends to homomorphic execution of typed functional programs, provided totality and normalization hold.

On the complexity side, we introduced the BNF Distinguishing Problem and argued for its hardness via a reduction from Graph Isomorphism, giving some justification for the cryptographic soundness of this approach.

Finally, we considered the substantial practical challenges and proposed a range of optimization techniques, from algorithmic improvements to hardware acceleration. While still far from a turnkey system, these ideas outline a roadmap from theoretical possibility to feasible implementation in niche scenarios.

Natural categories for future work include:

\begin{itemize}
  \item Formalizing the BNF Distinguishing Problem and providing stronger hardness results.
  \item Refining the polynomial functor and BNF constructions to find a sweet spot between expressivity and efficiency.
  \item Prototyping small-scale implementations to measure performance and guide further optimizations.
  \item Investigating extensions to handle controlled forms of effects or other logical frameworks.
\end{itemize}

\section*{Acknowledgements}  Thanks are due to Mike Stay for valuable comments and suggestions on an earlier version of this paper.

\section*{References:}
\begin{itemize}
  \item All, A. \& Assem, B. (2020). Homomorphic encryption schemes over polynomial rings.
  \item Avigad, J. et al. (2019). Quotients of BNFs and their properties.
  \item Boneh, D., Goh, E.-J., \& Nissim, K. (2005). Evaluating 2-DNF formulas on ciphertexts. TCC.
  \item Fontaine, C. \& Galand, F. (2007). A survey of homomorphic encryption for non-specialists. EURASIP Journal.
  \item Brady, E. (2013). Idris, a general-purpose dependently typed programming language: Design and implementation. Journal of functional programming, 23(5), 552-593.
  \item Hutton, G. (1999). A tutorial on the universality and expressiveness of fold. Journal of Functional Programming.
  \item Meredith, L. G., Goertzel, B., Warrell, J., \& Vandervorst, A. (2023). Meta-MeTTa: an operational semantics for MeTTa. arXiv preprint arXiv:2305.17218.
  \item Pfenning, F. (2009). Intuitionistic quantifiers and structural rules. Lecture notes.
  \item Tseng, J. H. et al. (2017). Homomorphic encryption of boolean formulas. Cryptology ePrint Archive.
  \item Furer, M. et al. (2020). Polynomial functors and BNFs: Theory and construction. Proceedings of CSL.\\
(Additional references to standard texts on type theory, category theory, and functional languages may be included as needed.)
\end{itemize}

\end{document}